\documentclass[12pt]{iopart}

\usepackage{iopams}

\expandafter\let\csname equation*\endcsname\relax

\expandafter\let\csname endequation*\endcsname\relax

\usepackage{amsmath}

\usepackage{graphicx}
\usepackage{float}
\usepackage{cancel}
\usepackage{cite}
\usepackage{hyperref}
\usepackage{subcaption} 
\usepackage{centernot}

\usepackage{calligra}
\DeclareMathAlphabet{\mathcalligra}{T1}{calligra}{m}{n}
\DeclareFontShape{T1}{calligra}{m}{n}{<->s*[2.2]callig15}{}
\newcommand{\scriptr}{\mathcalligra{r}\,}

\newcommand{\scriptt}{\mathcalligra{t}\,}
\begin{document}

\title[]{Semi-classical description of electrostatics and quantization of electric charge}

\author{Kolahal Bhattacharya}

\address{St. Xavier's College (Autonomous), Kolkata-700016, India}
\ead{kolahalbhattacharya@sxccal.edu}
\vspace{10pt}
\begin{indented}
\item[]
\end{indented}

\begin{abstract}

In this work, we present an explanation of the electric charge quantization based on a semi-classical model of electrostatic fields. We claim that in electrostatics, an electric charge must be equal to a rational multiple of the elementary charge of an electron. However, the charge is quantized if the system has certain boundary conditions that force the wavefunction representing an electric field to vanish at specific surfaces. Next, we develop the corresponding model for the electric displacement vector. It is demonstrated that a number of classical results, e.g. bending of field lines at the interface of two dielectric media, method of images, etc. are all consistent with the predictions of this model. We also present the possible form of Gauss's law (or Poisson's equation), to find the wavefunctions of the field from a source charge distribution, in this model. 

\end{abstract}


\section{Introduction}\label{Intro}
A prime example of the unresolved enigmas of theoretical physics is the quantization of electric charge. Millikan's famous oil drop experiment~\cite{millikan1913elementary} in 1909 demonstrated that electric charge always appears as an integral multiple of the elementary electric charge $e$. This result continues to hold even today, with great experimental accuracy~\cite{perl2009searches}. However, in spite of the extraordinary success of Maxwell's theory of electrodynamics and quantum electrodynamics (QED), the theoretical justification of this empirical result has remained unexplained until now. Moreover, the development of particle physics in the last seventy years has demonstrated that quarks, the constituent Fermions of Hadrons carry fractional charge $\pm e/3$, $\pm{2e}/3$ etc. There is no satisfactory explanation of these experimental results as well. In this paper, we will explain the charge quantization in the non-relativistic domain, as observed by Millikan.

In 1931, P. Dirac argued~\cite{dirac1931quantised} that the charge quantization must happen if there is a magnetic monopole. He showed that the unobservability of phase in the quantum domain allows for singularities, manifested as sources of magnetic fields. This leads to the condition that the product of electric and magnetic charges should be an integral multiple of $\hbar c/2$, where $\hbar$ represents the reduced Planck's constant and $c$ denotes the speed of light in the vacuum. This marked the beginning of an organised search for a magnetic monopole somewhere in the universe that continues even today because no magnetic monopole has ever been located\footnotemark[1]. \footnotetext[1]{Pierre de Maricourt of the thirteenth century tried to separate the poles of a magnet by breaking magnets into pieces~\cite{de1904magnet}. So, this search is perhaps a millennium old.} 

There are other thought-provoking motivations to search for magnetic monopoles. For example, the presence of a magnetic monopole will lead to a symmetric extension of Maxwell's laws for classical electrodynamics that are invariant under duality transformation~\cite{castellani2010dualities}. This symmetry will also suggest a generalised Lorentz force: 
\begin{equation} \vec{F}=q(\vec{\mathcal{E}}+\vec{v}\times\vec{\mathcal{B}})+g(\vec{\mathcal{B}}-\frac{\vec{v}}{c^2}\times\vec{\mathcal{E}})
\end{equation}
acting on a dyonic particle with electric charge $q$ and magnetic charge $g$ moving with a velocity $\Vec{v}$ in the electric field $\vec{\mathcal{E}}$ and magnetic field $\vec{\mathcal{B}}$. Now, if dyons exist, then space inversion is no longer a valid symmetry in electrodynamics, as noted by Ramsey~\cite{ramsey1958time}. Now, parity symmetry is broken maximally in weak interactions. This means that if magnetic monopoles exist, due to the quantization of electric charges, this will as well justify why parity is not a good symmetry of nature~\cite{cabibbo1962quantum, pintacuda1963magnetic, mcdonald2013poynting}. The magnetic monopole has also been proposed as a solution to the strong CP problem~\cite{zhang1994magnetic}. So, the issue of the quantization of electric charge is deeply connected to a large number of interesting questions in fundamental physics and therefore, there have been consolidated experimental efforts to search for magnetic monopoles. However, searches in the cosmic rays, bound matter, in colliders via direct and indirect ways never ever found the existence of monopoles~\cite{patrizii2015status}. The MoEDAL collaboration at CERN is an ongoing experiment and they recently reported the result of their first run~\cite{acharya2021first}, where they ruled out the existence of dyons carrying a magnetic charge up to five units of the Dirac charge and an electric charge up to 200 times the electron’s charge for dyons with a mass limits between 870 and 3120 GeV. Perhaps this hints that the charge quantization problem can be approached in a different route which does not require the existence of a magnetic monopole.

In this paper, we approach the problem using a quantum physical model of electrostatics. In the recent past, there have been some works in understanding the quantum or semi-classical nature of electrostatic fields~\cite{kay2022quantum,bhattacharya2021demystifying}. Both authors appear to agree that electrostatic fields could be described by non-travelling wavegroups, but otherwise, the frameworks are different. The second one among these two demystified the nonlocality problem of the Aharonov-Bohm effect which was a puzzle since 1959~\cite{aharonov1959significance}. After being proved experimentally~\cite{tonomura1986evidence}, this experiment was compared with the Michelson and Morley experiment of recent times~\cite{olariu1985quantum}. Apart from providing an explanation of the nonlocality problem, the article~\cite{bhattacharya2021demystifying} pointed out that electric charge must be a rational multiple of the elementary electric charge $e$. With this queue, the semi-classical model, introduced in~\cite{bhattacharya2021demystifying} will be used in our present quest in the hope that it can throw some light.  


We first obtain the prediction of the model in the context of the charge quantization problem. Then, we address another well-known piece of information that the electrostatic field lines, in some cases, could exhibit features very similar to the light rays in geometrical optics. For example, the bending of electric field lines at the interface between two dielectric media is analogous to the refraction of light. Similarly, in the examples of the method of images (which is an elegant method~\cite{jackson2007, greiner2012classical} to solve Laplace's equation for electrostatic potential under appropriate boundary conditions), the situations are very much similar to the reflection of light by mirrors. 

This model~\cite{bhattacharya2021demystifying} asserts that the electrostatic field can exhibit semi-classical nature when $e\Phi{t} \sim\hbar$ where $\Phi$ and $t$ represent electrostatic potential and time - over which a charge is subjected to the potential $\Phi$. If $e\Phi t\gg\hbar$, the classical nature of the fields manifests. On the other hand, if $e\Phi{t}\sim\hbar$, then in a source-free region, the wavefunctions of the electrostatic field satisfy a wave equation that has the form of a homogeneous Helmholtz equation. 
An analogous situation arises during the transition from ray optics to wave optics~\cite{gloge1969formal}. In the limit where the wavelength of the light cannot be neglected, the wave nature of light is manifested. Historically, Huygens contemplated plane and spherical wavelets of light, envelopes of which proceed in the forward direction for the propagation of light. In modern formalism, these waves are identified as the solutions to the reduced wave equation~\cite{torre2005linear}. Though this wave model of light lacks the polarisation picture, it can be used to explain the reflection and refraction of light. Therefore, the situations with optical analogies in electrostatics may indicate an underlying semi-classical model for classical electrostatics. Construction of such a mathematical model will be very exciting and it may reveal some unknown features of electrostatic field theory.


In~\cite{bhattacharya2021demystifying}, the wavefunctions of the electrostatic field have been introduced merely as mathematical objects representing fields in the regions devoid of source charges. In the current work, we extend this formalism in the presence of source charges in section~\ref{Section2} to find the equations corresponding to the wavefunctions that represent different components of the electrostatic field. In this formalism, we shall observe the presence of an anti-Hermitian operator that satisfies the spectral theorem and will investigate the classical limit of the model. Next, in section~\ref{Section3}, we present the proof of the quantization of charge, as observed in experiments. To address the question of the similarity between electrostatic field lines and light rays in geometrical optics, we shall develop the semi-classical description for the electric displacement vector $\vec{\mathcal{D}}$ in section~\ref{Section4}. In the following section~\ref{Section5}, we shall describe how the semi-classical model of $\vec{\mathcal E}$ and $\vec{\mathcal D}$ fields helps in understanding the refraction and reflection of field lines across a boundary of two media of different dielectric constants. In section~\ref{Section6}, we compare the relation of this formalism with the quantum limit of Gauss's law (or Poisson's equation), as presented in~\cite{kay2022quantum}. Finally, we will conclude with a discussion of the implication of these observations.
\section{Wave equation in Electrostatics}\label{Section2}
One can conceive a variational principle for the electrostatic field (and other curl-free vector fields)~\cite{bhattacharya2022unexplored}:
\begin{equation}\label{Eq1}
    \delta\int_{P_1}^{P_2}{\mathcal{E}}\ ds=0.
\end{equation}
In Eq.\eqref{Eq1}, the integral is evaluated along a curve that is always superimposed with the local direction of the field. We find that the field lines satisfy the Euler-Lagrange equation:
\begin{equation}\label{Eq2}
    \nabla{\mathcal{E}}=\frac{d}{ds}\left(\mathcal{E}\frac{d{\bf r}}{ds}\right),
\end{equation}
exactly in a way similar to the light rays~\cite{lakshminarayanan2002lagrangian}. It has recently been shown that the electrostatic field may exhibit a semi-classical behaviour if $e\Phi t\sim\hbar $~\cite{bhattacharya2021demystifying}. Under source-free conditions, one can define the electric field operator as a momentum conjugate to the position coordinates:
\begin{align}\label{Eq3}
    \hat{\vec{p}}\psi_E=-i\bar\gamma\nabla\psi_E=(\hat{\bf x} \hat{p}_x+ \hat{\bf y} \hat{p}_y+\hat{\bf z} \hat{p}_z)\psi_E=(\hat{\bf x}\mathcal{E}_x+\hat{\bf y}\mathcal{E}_y+\hat{\bf z}\mathcal{E}_z)\psi_E=\vec{\mathcal{E}}\psi_E,
\end{align}
where ($\hat{p}_x,\hat{p}_y,\hat{p}_z$) denotes the momentum operators in $(x,y,z)$ directions along which $\hat{\bf x},\hat{\bf y},\hat{\bf z}$ are the unit vectors; $\bar \gamma\equiv\gamma/(2\pi)=\hbar/(e\cdot t)$ is a scaling factor, which represents $1/(2 \pi)$ times the minimum possible electrostatic potential $\gamma$ in the problem. In a region devoid of source charge density $\rho$ (where the conjugate momentum operators $\hat {p}_x$ etc. do not operate on the corresponding components of the field e.g. $\mathcal{E}_x$ etc.), the non-travelling wavefunction $\psi_E$ satisfies:
\begin{equation}\label{Eq4}    \bar\gamma^2\nabla^2\psi_E+\mathcal{E}^2\psi_E=0.
\end{equation}
We can readily verify that $\psi_E={\bf e}^{i{e\Phi t}/{\hbar}}$ is a solution to this equation. The presence of the variable time factor ($t$) in $\psi_E$ may appear contradictory in the context of standard electrostatics problems, where one is interested in the time-averaged electric field or potential at a test point, due to some given charge distribution. We will find that the classical results follow from the boundary conditions on $\psi_E$ but they do not depend on $\psi_E$ itself. In certain special circumstances, when we need to consider the electromagnetic communication between two bodies, the time factor can be expressed as $L/c$ where $L$ is the distance between the bodies, and $c$ is the speed of light in free space.

The wave equation~\eqref{Eq4} is valid in the region of space devoid of source charge. However, in the presence of source charge density $\rho\neq{0}$, the conjugate momenta $\hat{p}_x$, etc. can operate on the components of the field e.g. $\mathcal{E}_x$, etc. Not only that, the distribution of charges can be different in different directions. For example, along the axis of the charged plate capacitors, there is a non-zero divergence of the electric field at the capacitor plates, due to the presence of source charge. However, in the directions perpendicular to the axis, the field does not have divergence. For such situations, it is more meaningful to contemplate different wavefunctions of the electrostatic field along different directions:
\begin{subequations}
\begin{align}
-i\bar\gamma\frac{\partial\psi_{E_x}}{\partial x}&=\mathcal{E}_x\psi_{E_x}\label{Eq5a}\\
-i\bar\gamma\frac{\partial\psi_{E_y}}{\partial y}&=\mathcal{E}_y\psi_{E_y}\label{Eq5b}\\
-i\bar\gamma\frac{\partial\psi_{E_z}}{\partial z}&=\mathcal{E}_z\psi_{E_z}\label{Eq5c}
\end{align}
\end{subequations}
In a medium of permittivity $\epsilon_0$, wave equation for wavefunction $\psi_{E_z}$ in $z$ direction take the form:
\begin{subequations}
\begin{align}
    -i\bar\gamma\frac{\partial^2\psi_{E_z}}{\partial z^2} &=\frac{\partial\mathcal{E}_z} {\partial z}\psi_{E_z}+\mathcal{E}_z\frac{\partial\psi_{E_z}}{\partial z}\nonumber\\
    &=\frac{\rho_z}{\epsilon_0}\psi_{E_z}+\mathcal{E}_z\frac{\mathcal{E}_z\psi_{E_z}}{-i\bar\gamma}\nonumber\\
\implies&\bar\gamma^2\frac{\partial^2\psi_{E_z}}{\partial z^2}+\mathcal{E}_z^2 \psi_{E_z}=i\bar\gamma\frac{\rho_z}{\epsilon_0}\psi_{E_z}\label{Eq6a}\\
\implies&\left[-\frac{\bar\gamma^2}{2\left(\frac{1}{\epsilon_0}\right)}\frac{\partial^2}{\partial z^2}-\frac {\epsilon_0\mathcal{E}_z^2}{2}\right]\psi_{E_z}=i\left (-\frac{1}{2}\rho_z\bar\gamma\right)\psi_{E_z}\label{Eq6}
\end{align}
\end{subequations}
Here, we have denoted ${\partial{\mathcal{E}_z}}/{\partial z}=\rho _z$. For $x$ and $y$ components, it is possible to have corresponding equations. Eq.\eqref{Eq6} has the form of the time-independent Schr$\rm{\ddot o}$dinger's equation. The first and second terms on the left-hand side denote the kinetic and potential energy density. We notice that the inverse of the permittivity plays the role of mass of the $\psi_{E_z}$ field. The (imaginary) energy density is given by the factor $-i\rho_z\bar\gamma/2$ on the right-hand side of Eq.\eqref{Eq6}. However, unlike Schr$\rm{\ddot o}$dinger's equation, it cannot be interpreted as an equation describing the evolution of a wavefunction in an external potential barrier. Rather, this equation relates the electric field component and the corresponding wavefunction, to the related source charge distribution.

The presence of $i$ on the right-hand side of Eq.\eqref{Eq6} shows that the operator on the left-hand side, acting on $\psi_{E_z}$ is an anti-Hermitian operator (with imaginary eigenvalues), unlike the standard linear Hermitian operators associated with physical observables. In fact, it is similar to the complex scalar field operators in quantum field theory. From a mathematical point of view, it is a normal operator. Such an operator $\hat N$ is defined on a complex vector space $\mathcal{H}$, such that it commutes with its Hermitian adjoint, i.e.: $\hat N\hat N^{\dagger}=\hat N^{\dagger}\hat N$. Both Hermitian, as well as anti-Hermitian operators, are examples of normal operators that assume the form of a diagonal matrix with respect to an orthonormal basis, in accordance with the spectral theorem. However, concrete examples of the latter are commonly not found. Naturally, the full potential of these operators has not been fully explored in physics discourses. There have been some pioneering works by Bender~\cite{bender2005non} on the possible use of non-Hermitian operators in quantum field theory. The idea has also been supported by R Penrose on p. 539 of `Road to Reality'~\cite{penrose2004road}. In this paper, we will find tangible examples of these operators in the context of electrostatic field theory.

It may be interesting to explore the physical meaning of the imaginary energy density in Eq.\eqref{Eq6}. Zhang~\cite{zhang2008electric} conjectured the electric charge as an imaginary form of energy. Using this, he showed a pathway of unification of gravitational and electrical forces classically. Similar or related ideas have been expressed by other authors~\cite{bhattacharya2023charge}. However, these ideas appear to be more speculative. It seems that there may be some connection, but it will be premature to say that one implies the other. In fact, it seems that the sense in which they defined `imaginary energy' of a charged particle is somewhat different than the sense in which imaginary energy density appears in the current framework. About the nature of the solutions, we comment that in the absence of source charge, for which the right-hand-side of Eq.\eqref{Eq6} vanishes, the equation is just a simple harmonic oscillator equation with a position-dependent frequency, as pointed out in~\cite{bhattacharya2021demystifying}. But in the presence of source charge, the actual solution must be worked out using boundary conditions.

Before we go forward, it is worthwhile to investigate the combined three-dimensional version of the problem. One way to accomplish that is by adding Eq.\eqref{Eq6} with its $x$ and $y$ counterparts. However, this does not give new mathematical insight into the system. On the contrary, if we choose to represent the components of the electrostatic field ($\mathcal{E}_z$ etc.) in terms of the partial derivatives of the logarithms of the corresponding wavefunctions ($\psi_{E_z}$ etc.) on the basis of Eq.\eqref{Eq5c} etc. then by adding all the component equations, we can deduce an interesting result:
\begin{align}\label{EqQuantumGauss}
    &-i\bar\gamma\left(\hat{\bf x}\frac{\partial\ln\psi_{E_x}}{\partial x}+\hat{\bf y}\frac{\partial\ln\psi_{E_y}}{\partial y}+\hat{\bf z}\frac{\partial\ln\psi_{E_z}}{\partial z}\right)=(\hat{\bf x}\mathcal{E}_x+\hat{\bf y}\mathcal{E}_y+\hat{\bf z}\mathcal{E}_z)\nonumber\\
&\implies-i\bar\gamma\left(\hat{\bf x}\frac{\partial}{\partial x}+\hat{\bf y}\frac{\partial}{\partial y}+\hat{\bf z}\frac{\partial}{\partial z}\right)\cdot\left(\hat{\bf x}\frac{\partial\ln\psi_{E_x}}{\partial x}+\hat{\bf y}\frac{\partial\ln\psi_{E_y}}{\partial y}+\hat{\bf z}\frac{\partial\ln\psi_{E_z}}{\partial z}\right)=\nabla\cdot\vec{\mathcal{E}}=\frac {\rho}{\epsilon_0}\nonumber\\
&\implies-i\bar\gamma\left(\frac{\partial^2\ln\psi_{E_x}}{\partial x^2}+\frac{\partial^2\ln\psi_{E_y}}{\partial y^2}+\frac {\partial^2\ln\psi_{E_z}}{\partial z^2}\right)=\frac {\rho}{\epsilon_0}\nonumber\\
&\implies-i\bar\gamma\left(\hat{\bf x}\frac{\partial^2}{\partial x^2}+\hat{\bf y}\frac{\partial^2}{\partial y^2}+\hat{\bf z}\frac{\partial^2}{\partial z^2}\right)\cdot(\hat{\bf x}\ln\psi_{E_x}+\hat{\bf y}\ln\psi_{E_y}+\hat{\bf z}\ln\psi_{E_z})=\frac{\rho}{\epsilon_0}
\end{align}
This equation demonstrates that in the vicinity of non-zero source charge, one must talk about a vector of wavefunctions $(\hat{\bf x}\ln\psi_{E_x}+\hat{\bf y}\ln\psi_{E_y}+\hat{\bf z}\ln\psi_{E_z})$ in three-dimensional space, instead of $\psi_E$ which was used in the absence of source charge. If there is a spherical symmetry, then Eq.\eqref{EqQuantumGauss} will assume a simpler form. We are tempted to conjecture that this equation can be regarded as the quantum mechanical version of Gauss's law (or Poisson's equation). It must be noted that we had to introduce a three-dimensional vector of wavefunctions and a new vector differential operator:
\begin{equation}\label{Eq7}
\left(\hat{\bf x}\frac{\partial^2}{\partial x^2}+\hat{\bf y}\frac{\partial^2}{\partial y^2}+\hat{\bf z}\frac{\partial^2}{\partial z^2}\right)
\end{equation}
that operates on it. The current author is not aware of any examples where such an operator has been applied.\\
Now, let us note that the complex conjugation of Eq.\eqref{Eq6a} gives:
\begin{equation}\label{Eq8}
    \bar\gamma^2\frac{\partial^2\psi_{E_z}^*}{\partial z^2}+ \mathcal{E}_z^2\psi_{E_z}^*=-i\bar\gamma\frac{\rho_z}{\epsilon}\psi_{E_z}^*.
\end{equation}
Multiplying Eq.\eqref{Eq6a} by $\psi_{E_z}^*$ and Eq.\eqref{Eq8} by $\psi_{E_z}$, then adding the resulting two equations, we get
\begin{align}\label{Eq9}
    &\bar\gamma^2\left(\psi_{E_z}^*\frac{\partial^2\psi_{E_z}}{\partial z^2}+\psi_{E_z}\frac{\partial^2\psi_{E_z}^*}{\partial z^2}\right)+2\mathcal{E}_z^2\psi_{E_z}^*\psi_{E_z} =0\nonumber\\ \implies&\bar\gamma^2\left(\psi_{E_z}^*\frac{\partial^2\psi_{E_z}}{\partial z^2}+\psi_{E_z}\frac{\partial^2\psi_{E_z}^*}{\partial z^2}\right)+2\left(i\bar\gamma\frac{\partial\psi_{E_z}^*}{\partial z}\right)\cdot\left(-i\bar\gamma \frac{\partial\psi_{E_z}}{\partial z}\right)=0\nonumber\\
    \implies&\bar\gamma^2\left(\psi_{E_z}^*\frac{\partial^2\psi_{E_z}}{\partial z^2}+\psi_{E_z}\frac{\partial^2\psi_{E_z}^*}{\partial z^2}\right)+ 2\bar\gamma^2\frac{\partial\psi_{E_z}^*}{\partial z}\frac{\partial\psi_{E_z}}{\partial z}=0\nonumber\\
    \implies&\frac{\partial^2}{\partial z^2}\left|\psi_{E_z} \right|^2\equiv\frac{\partial^2}{\partial z^2}\left(\psi_{E_z}^*\psi_{E_z}\right)=0,
\end{align}
where, in the second equality, we used Eq.\eqref{Eq5c} and its complex conjugate. One can easily verify that in the absence of the source charge distribution $\rho$, the corresponding equation becomes:
\begin{equation}\label{Eq10}
 \nabla^2|\psi_E|^2=0,   
\end{equation}
which can perhaps be anticipated. As such, $\psi_E$ has a spherical symmetry and we do not need to distinguish between the directions in the source-free case. Invoking Born's probability interpretation, we find that in the source-free region, the probability density of finding $\psi_E$ at a point in space is a Harmonic function (solution of Laplace's equation), from Eq.\eqref{Eq10}. In the presence of a source charge, we must be concerned with finding the probability density of wavefunction of the electrostatic field in a given direction. That still remains a Harmonic function, in that direction. Perhaps this suggests a possible connection between the modulus square of the wavefunction of the electric field and the electrostatic potential. If the boundary conditions on the potential are the same as those on the wavefunction, (say, if both of them are equal to zero on a boundary), then the uniqueness theorem states that the modulus squared wavefunction is the same as the potential. In that case, one can find the electrostatic potential, by solving Eq.\eqref{Eq6}. We comment that if only one component (say, $z$ component) of the curl of a vector field is zero (or equivalently if the closed-loop line integral for that component of the vector field is zero), then it is possible to find the semi-classical model for that component.

We note that the Eq.\eqref{Eq4} admits both positive and negative signed exponents as basis wavefunctions, i.e. $ \psi_E={\bf e}^{\pm i{q\Phi t}/{\hbar}}$. If $q=-e$, the normalised solution of Eq.\eqref{Eq4} will be $\Psi _E=\int u(\vec{\mathcal E}){\bf e}^{\pm i{e\Phi t}/{\hbar}}d\vec{\mathcal E}$ where $\Phi=-\int{\vec {\mathcal E}}\cdot d{\bf r}$. However, Eq.\eqref{Eq6} admits only $\psi_{E_z}={\bf e}^{-i{q\int(\mathcal{E}_z dz)t}/{\hbar}}$ as the basis wavefunctions. For $q=-e$, the corresponding normalizable solutions to Eq.\eqref{Eq6} will be $\psi_{E_z}=\int u({\mathcal E}_z){\bf e}^{i{e(\int {\mathcal E}_z dz) t}/{\hbar}}d{\mathcal{E}_z}$. 
Before we conclude this section, let us check the classical limit of Eq.\eqref{Eq6}. In the investigation of the classical limit of quantum mechanics, a standard approach is to write the wavefunction as $\psi=A{\bf e}^ {iS/\hbar}$ where $A$ and $S$ are real quantities. If we substitute this into the Schr$\rm{\ddot o}$dinger's equation and take the limit, we find that $S$ (classical action) satisfies the Hamilton-Jacobi equation. 

Now, in the context of Eq.\eqref{Eq6}, we can proceed in a similar manner. Let us write the wavefunction as $\psi_{E_z}=A{\bf e}^ {-i\frac{\Phi}{\bar\gamma}}$, noting that the electrostatic potential plays a role similar to the classical action. Substituting this function into Eq.\eqref{Eq6}, we find:
\begin{subequations}
    \begin{align}
    \bar\gamma^2\frac{\partial^2\psi_{E_z}}{\partial z^2}&=\bar\gamma^2\frac{d^2A}{dz^2}\left({\bf e}^{-i\frac{\Phi}{\bar \gamma}}\right)-2i\bar\gamma\frac{dA}{dz} \frac{\partial\Phi}{\partial z}\left( {\bf e}^{-i\frac{\Phi}{\bar\gamma}} \right)-i\bar\gamma\frac{\partial^2\Phi}{\partial z^2}\left(A{\bf e}^{-i\frac {\Phi}{\bar\gamma}}\right)-\mathcal{E}_z ^2\left(A{\bf e}^{-i\frac{\Phi}{\bar \gamma}}\right)\label{Eq13a}\\
    &=-\mathcal{E}_z^2\left(A{\bf e}^ {-i\frac{\Phi}{\bar\gamma}}\right)+i\bar \gamma\frac{\rho_z}{\epsilon_0}\left(A{\bf e}^{-i\frac{\Phi}{\bar\gamma}}\right)\label{Eq13b}
\end{align}
\end{subequations}
If we take the limit $\bar\gamma\rightarrow0$ in Eq.\eqref{Eq13a} and Eq.\eqref{Eq13b}, then we have:
\begin{align}
\lim_{\bar\gamma\to 0} -i\bar\gamma \frac{\partial^2\Phi}{\partial z^2}\left(A{\bf e}^{-i\frac{\Phi}{\bar\gamma}}\right)&=\lim_{\bar\gamma\to 0}i\bar\gamma\frac{\rho_z}{\epsilon_0}\left(A{\bf e}^{-i\frac{\Phi}{\bar\gamma}}\right)\nonumber\\
\implies \frac{\partial^2\Phi}{\partial z^2}&=-\frac{\rho_z}{\epsilon_0}
\end{align}
This is Gauss's law in one dimension. If the same is done in all directions, we get the traditional differential form of Gauss's law in three dimensions.

\section{Proof of quantization of electric charge}\label{Section3}
In~\cite{bhattacharya2021demystifying}, it has been argued that the charge $q$ in an electrostatic system, in general, should be a rational multiple of the elementary electric charge. This can be proved in the following way. Consider a normalizable wavefunction $\Psi_E$ in regions with non-zero field value. We demand that $\Psi_E$ must remain the same for a constant change in potential, just the way the classical electric field remains unaffected by a constant change in electrostatic potential. Referring to the form of $\Psi_E$, we notice that the Fourier coefficient $u(\vec{\mathcal{E}})$ will remain invariant under the transformation $\Phi \rightarrow\Phi+\Phi_0$ where $\Phi_0$ is a constant. This leads to:
\begin{equation}\label{Eq11}
    \int u(-\nabla\Phi){\bf e}^{i\frac{q\Phi t}{\hbar}} d\vec{\mathcal{E}}=\int u(-\nabla(\Phi+\Phi_0)){\bf e}^{i\frac{q(\Phi+\Phi_0)t}{\hbar}}d\vec{\mathcal{E}}\\
    \implies {\bf e}^{i\frac{q\Phi t}{\hbar}}={\bf e}^{i\frac{q(\Phi+\Phi_0) t}{\hbar}}
\end{equation}
In Eq.\eqref{Eq11}, we can equate the integrands, because the integrals are equal for arbitrary boundaries (any pair of upper and lower limits) and a constant $\Phi_0$. Now, for an integer $n \in\mathbf{N}$, this leads to the following condition:
\begin{align}
    \frac{q\Phi t}{\hbar}&=2n\pi+\frac{q(\Phi+\Phi_0)t}{\hbar}\nonumber\\
\implies q\Phi_0t&=-2n\pi\cdot\hbar=-nh\nonumber\\
\implies q&=-\frac{nh}{\Phi_0t}=-\frac{n}{\Phi_0}(\gamma\cdot e)=-\frac{n}{(\Phi_0/\gamma)}e=-\frac{n}{N}e,
\end{align}
where in the last equation, we have used $\Phi_0=N \gamma$, where $N$ is an integer. The potential $\Phi_0$ represents an area in the phase space constituted by coordinates and conjugate momenta (electric field). This area must be an integral multiple of the unit (minimum) potential $\gamma$. Since $\Phi_0$ can be positive as well as negative, so $N$ can also be both positive and negative.
This result shows that the charge should be a rational multiple of $e$, but does not explain why charges should be quantized. In the following, we provide more direct proof of the said quantization.  Let us consider the problem in one dimension (say, in $z$ direction). 
If some charge is distributed on a conductor, e.g. to a plate of a parallel plate capacitor located at $z=a$ (whose another plate at $z=0$ is grounded), then the charge density is given by $\rho=\sigma \delta_D(z-a)$, where $\delta_D$ represents Dirac delta function. Then, from Eq.\eqref{Eq6}, we can say:
\begin{align}\label{Eq12}
-\frac{\hbar^2}{2\left(\frac{e^2 t^2}{\epsilon_0}\right)}\frac{d^2\psi_{E_z}}{dz^2}-\frac{\epsilon_0}{2}\mathcal{E}_z^2\psi_{E_z}&=-i\bar\gamma\frac{\sigma}{2}\delta_D(z-a)\psi_{E_z}
\end{align}
Let us check the boundary conditions of $\psi_{E_z}$ and $\mathcal{E}_z$. Between the plates ($0<z<a$), $\mathcal{E}_z\neq0$, and $\psi_{E_z}=\int a(\mathcal{E}_z){\bf e}^{i{e({\int\mathcal E}_z dz)t}/{\hbar}} d{\mathcal{E}_z}$. Exactly on the surface of the conductor $z=a$, $\psi_{E_z}=0$, otherwise the RHS of Eq.\eqref{Eq12} will diverge. Assuming that the electric field between the plates is a constant $\mathcal{E}_0$,
we are led to the condition that $\sin\left({e\mathcal E_0a t}/\hbar\right)=0$ (or $\cos\left({e\mathcal E_0a t}/\hbar \right)=0$). If we adopt the $\sin$, we get:
\begin{align}\label{Eq13}
    \left(\frac{e\mathcal E_0a t}{\hbar}\right)=n\pi\nonumber\\
\implies\mathcal E_0=n\frac{\pi\hbar}{ea t}=n\pi\frac {\bar\gamma}{a}
\end{align}
Note that $\bar\gamma$ has the unit of electrostatic potential. Since the electric field just outside the conductor is related to the charge density by $\sigma =\epsilon_0\mathcal E_0$, it follows that the original charge $Q(=\sigma A)$, given to the plate of area $A$ at $z=a$, must be quantized as well. If we adopt $\cos$, even then the electric field - and the source charge of that field - would be quantized. 

There is no loss of generality in selecting a specific configuration of conductors. We could as well choose a charged conductor of arbitrary shape. Choose a point $P$ on the surface and call the outward normal direction the $z$ direction. The value of the field just outside $P$ at a distance $\Delta$ ($\rightarrow0$) from the surface, would still be $\mathcal{E}_z=\sigma/\epsilon _0$. The integral $\int\mathcal{E}_z dz$ will reduce to $\mathcal{E}_z \Delta$. This does not change any of the arguments used in Eq.\eqref{Eq13}. We discuss the case of quantization of charge in dielectric systems in the next section.

The eigenvalue at the right-hand side of Eq.\eqref{Eq6} i.e. $-i{\rho_z\bar\gamma}/2$ must be quantized according to appropriate boundary conditions, exactly in the same way the energy eigenvalues of a particle in a potential well are quantized. This way, the boundary condition requires the quantization of the charge in the source distribution. This description, therefore, removes the need for magnetic monopoles due to the quantization of electric charge.

The preceding discussion suggests that the quantization of electric charge has two aspects. Fundamentally, it is not a quantized entity. In fact, it is a rational multiple of the elementary electric charge $e$, as far as electrostatics is concerned. However, it becomes quantized, when the boundary condition requires it to be so (through Eq.\eqref{Eq6}). 
\section{Electric Displacement vector $\vec{\mathcal D}$}\label{Section4}
If we assume that the medium is a linear dielectric with the polarisation vector $\vec{\mathcal{P}}=\epsilon_0 \chi_e\vec{\mathcal{E}}$, where $\chi_e$ represents the electric susceptibility, then the electric displacement is given by $\vec{\mathcal D}=\epsilon\vec{\mathcal{E}}= \epsilon_0(1+\chi_e)\vec{\mathcal E}$. In general, $\nabla\times\vec{\mathcal D} =\nabla\times\vec{\mathcal{P}}$. Now, if in a given problem,
$\nabla\times\vec{\mathcal{P}}={\bf0}$ (or, equivalently, $\oint \vec{\mathcal{P}}\cdot d\vec{l}=0$), the dielectric potential $\Phi_D$ can be expressed in terms of free charge density $\rho_f$ as:
\begin{equation*}
    \Phi_D=\epsilon\Phi=\frac{1}{4\pi}\int\frac{\rho_f}{|{\bf r}-{\bf r'}|}d\tau'\hspace{1.0cm}({\rm if}\hspace{0.25 cm}\nabla \times\vec{\mathcal{P}}={\bf0}).
\end{equation*}
Based on~\cite{bhattacharya2022unexplored}, in this case also one can conceive $\delta\int\mathcal D ds=0$, where this integral is evaluated along a curve, always superimposed with the local direction of $\vec{\mathcal{D}}$. This can be used to obtain a semi-classical model for the field $\vec{\mathcal D}$ in the limited cases where $\nabla \times\vec{\mathcal P}={\bf0}$. The eigenvalue equation for $\vec{\mathcal{D}}$ field should be given by:
\begin{equation}\label{Eq14}
    -i\frac{\gamma_D}{2\pi}\nabla\psi_D=\vec{\mathcal{D}}\psi_D,
\end{equation}
where ${\gamma_D}/(2\pi)(=\bar\gamma_D$, say) can be determined in the following way: the change in the action of a charge $q$ introduced in a medium of permittivity $\epsilon(\neq\epsilon_0)$, if it is subjected to potential $\Phi=\Phi_D/\epsilon$ is $\Delta S=-q\Phi t=-q(\Phi_D/\epsilon)t$. We define $\bar\gamma_D$ as the minimum value of $\Phi_D$ corresponding to the minimum action $\hbar$. If we take $q=-e$, then $\bar\gamma_D={\epsilon\hbar}/({e\cdot t})=\epsilon\bar\gamma$. 

If only one (say, $z$) component of the electric displacement vector is irrotational i.e. $\left({\partial\mathcal{D}_y}/{\partial x}-{\partial \mathcal{D}_x}/{\partial y}\right)=0$ [or, equivalently $\oint \mathcal{D}_z dl=0$ along a chosen closed contour], then we can get a semi-classical model with wavefunction $\psi_{D_z}$ that represents only the $z$ component of $\vec{\mathcal D}$:
\begin{subequations}
\begin{align}
    -i\frac{\gamma_D}{2\pi}\frac{\partial\psi_{D_z}}{\partial z}&=\mathcal{D}_z\psi_{D_z}\label{Eq14c}\\
\implies\bar\gamma_D^2\frac{\partial^2}{\partial z^2}\psi_{D_z}+{\mathcal{D}_z}^2&\psi_{D_z}=i\bar\gamma_D\rho_f\psi_{D_z}\label{Eq15b}.
\end{align}    
\end{subequations}
Using the value of $\bar\gamma_D$, Eq.\eqref{Eq15b} (for linear dielectrics) reduces to:
\begin{align}\label{Eq16}
\bar\gamma^2\frac{\partial^2\psi_{D_z}}{\partial z^2}+{\mathcal{E}_z}^2 \psi_{D_z}=i\frac{\rho_f}{\epsilon}\bar \gamma\psi_{D_z}\nonumber\\
\implies\left[-\frac{\bar\gamma^2}{2\left(\frac{1}{\epsilon}\right)}\frac{\partial^2}{\partial z^2}-\frac{{\mathcal{D}_z}{\mathcal{E}_z}}{2}\right]\psi_{D_z}&=i\left(-\frac{1}{2}\rho_f\bar\gamma\right)\psi_{D_z}.
\end{align}
The normalised solution of Eq.\eqref{Eq16} can be constructed as $\Psi_{D_z}=\int v({\mathcal{D}_z}){\bf e}^{-i{\Phi_{D_z}}/{\bar\gamma_D}}d{\mathcal D_z}=\int v({\mathcal D_z}){\bf e}^{-i{\Phi_z}/{\bar\gamma}}d{{\mathcal{D}_z}}$. Comparing with the form of the solution to Eq.\eqref{Eq6} (discussed just before section~\ref{Section3}), we find that the plane wave basis of the semi-classical wavefunction of the fields $\vec {\mathcal E}$ and $\vec{\mathcal{D}}$ are identical. But the coefficients are different, as expected.

In principle, one can always do the same exercise for the vector field $\vec{\mathcal{D}}-\vec{\mathcal{P}}$ (=$\epsilon_o\vec{\mathcal{E}}$), whose curl is zero. Not surprisingly, one finds the total charge density as the sum of free charge density $\rho_f$ and bound charge density $\rho_b$ (from $-\nabla\cdot\vec{\mathcal{P}}$ term).

We make the following observation about the quantization of electric charge in the case of dielectrics. Unlike the conductors, here we may have free and bound charges and they might reside within the body as well as on the interface. In addition, the electrostatic field and the displacement vectors do not remain perpendicular to the interface. But the second point does not pose a serious problem, because typically we would still deal with the same basis wavefunctions in a given direction, as found just before (we shall see an example in the context of dielectric half-plane image problem). So, for surface charge distribution, which can be represented by a delta function, one can predict the existence of quantized charges. That conclusion will not hold for the smooth continuous volume charge distributions. Most likely, such configurations will not have quantized charges. However, if the volume charge is made up of many individual point charges embedded in the medium, then the source term is composed of a summation over delta functions located at those points. These point charges must then be quantized.

\section{Electrostatic refraction}\label{Section5}
Let us consider two halves of the full space filled with linear dielectric materials with dielectric constants $\epsilon_1$ and $\epsilon_2$. We consider the oblique incidence of the electric field line at the boundary (see the following Figure~\ref{fig2}). We would like to approach this problem from the semi-classical description of the fields. To accomplish that, we make several observations:\\
\begin{figure}[ht]
\centering
\begin{subfigure}{.35\textwidth}
  \centering
  \captionsetup{justification=centering}
  \includegraphics[height=6.0 cm, width= 7.5 cm]{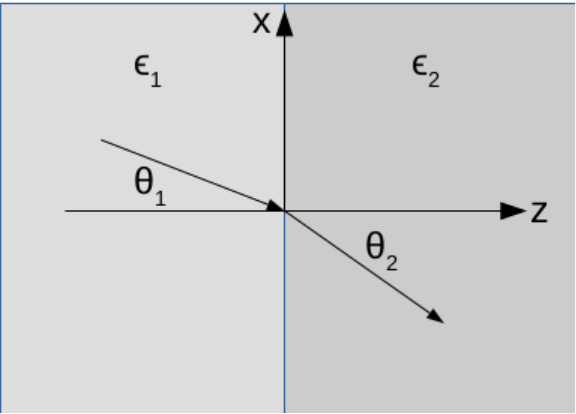}
  \caption{}
  \label{fig2a}
\end{subfigure}
\hspace{1.5 cm}
\begin{subfigure}{.45\textwidth}
  \centering
  \captionsetup{justification=centering}
  \includegraphics[height=6.0 cm, width= 7.5 cm]{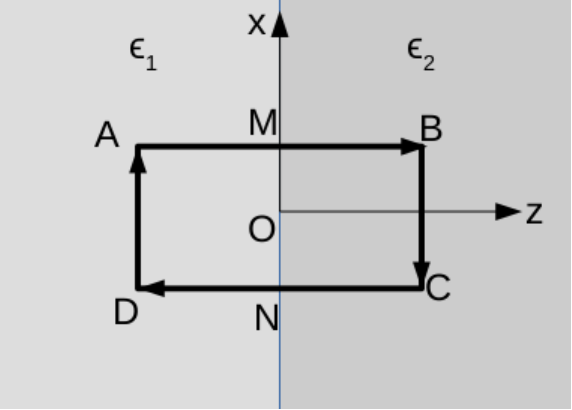}
  \caption{}
  \label{fig2b}
\end{subfigure}
\caption{(a) Direction change of electric field lines at the interface between two dielectric media, (b) closed loop line integral of $\vec{\mathcal{P}}$ vector on both sides of the interface.
}
\label{fig2}
\end{figure}
(a) On the interface, the free charge density $\rho_f=0$, but the total charge density $\rho$ which is defined as the sum of free charge density and the bound charge density, is not zero. In this context, boundary conditions are written as:
\begin{subequations}
\begin{align}
    \bar\gamma^2\frac{d^2}{dx^2}\psi_{E_x}+\mathcal{E}_x^2\psi_{E_x}&=0\label{Eq17a}\\
    \bar\gamma^2\frac{d^2}{dy^2}\psi_{E_y}+\mathcal{E}_y^2\psi_{E_y}&=0\label{Eq17b}\\
    \bar\gamma^2\frac{d^2}{dz^2}\psi_{E_z}+\mathcal{E}_z^2\psi_{E_z}=i\bar\gamma\frac{\sigma_b}{\epsilon_0}&\delta_D(z)\psi_{E_z}\label{Eq17c},
\end{align}
\end{subequations}
where we denoted the wavefunctions corresponding to individual components of the electric field with separate subscripts. Referring to Eq.\eqref{Eq17c}, we note that the existence of the delta function at $z=0$ implies discontinuity in the first $z$ derivative of $\psi_{E_z}$, which (from Eq.\eqref{Eq5c}) also implies discontinuity in the $z$ component of the electric field. However, along the tangential direction, the field is continuous, since there is no infinite jump in these directions. So, one has $\mathcal{E}_{1x,y}=\mathcal{E}_{2x,y}$ at the interface.

(b) At the interface, we can show that  $\oint\mathcal{P}_zdl=0$ (which also implies $\oint\mathcal{D}_zdl=0$) along a closed rectangular contour, going into and turning back from both sides of the interface, as can be seen with reference to Fig.~\ref{fig2b}:
\begin{align*}
\oint\mathcal P_zdl&=\int_A^M P_zdl+\int_M^B P_zdl+\cancelto{0}{\int_B^C P_zdl}+\int_C^N P_zdl+\int_N^D P_zdl+\cancelto{0}{\int_D^A P_z dl}\\
&=\epsilon_0\left(\chi_e^{(1)}\mathcal{E}_z^{1}\cdot AM+ \chi_e^{(2)}\mathcal{E}_z^{2}\cdot MB- \chi_e^{(2)}\mathcal{E}_z^{2}\cdot CN- \chi_e^{(1)}\mathcal{E}_z^{1}\cdot ND\right)=0.
\end{align*}
This allows us to develop the semi-classical description for $\mathcal{P}_z$ and hence for $\mathcal{D}_z$. Since $\rho_f=0$ on the interface, in this case, Eq.\eqref{Eq15b} has a simple form:
\begin{equation}\label{Eq18}
    \bar\gamma_D^2\frac{d^2}{dz^2}\psi_{D_z}+\mathcal{D}_z^2\psi_{D_z}=0.
\end{equation}
As before, we can argue in favour of continuity of $\mathcal{D} _z$ across the interface due to the absence of infinite jump in the first derivative of $\psi_D$. So, we have $D_{1z}=D_{2z}$. Using $\vec{\mathcal{D}}=\epsilon\vec{\mathcal{E}}$, and the continuity of the tangential component of the electric field, we deduce the relation:
\begin{equation}\label{Eq19}
    \frac{\epsilon_1}{\epsilon_2}=\frac{\tan\theta_1}{\tan\theta_2}.
\end{equation}
Finally, we evaluate $\oint\mathcal P_xdl=0$ along the same contour for completeness:
\begin{align*}
\oint\mathcal P_xdl&=\cancelto{0}{\int_A^M P_xdl}+\cancelto{0}{\int_M^B P_xdl}+\int_B^C P_xdl+ \cancelto{0}{\int_C^N P_xdl}+\cancelto{0}{\int_N^D P_xdl}+\int_D^A P_x dl\\
&=\epsilon_0\left(-\chi_e^{(2)}\mathcal{E}_z^{2}\cdot BC+ \chi_e^{(1)}\mathcal{E}_z^{1}\cdot DA\right)\neq0.
\end{align*}
Thus, the corresponding semi-classical method is not possible for the tangential components $\vec{\mathcal{P}}_{x,y}$ (and for $\vec{\mathcal{D}}_{x,y}$).
\subsubsection{Dielectric half-plane image problem}
Let us now consider the Dielectric half-plane image problem which is covered in standard texts~\cite{greiner2012classical, jackson2007}. The electric field $\vec{\mathcal{E}}_1$ originates from a charge $q$ located at a distance $-z_0$ in the medium with dielectric constant $\epsilon_1$. The problem seeks to find the potential function that will satisfy the boundary conditions. The standard approach to the problem is to assume that there are two image charges. One of them ($q'$) is located within the medium with dielectric constant $\epsilon_2$ at a distance $z_0$ from the boundary and the other ($q''$) is placed at the same location as the original charge $q$. Solving the boundary conditions, the values of the image charges are:
\begin{equation}\label{Eq20}
    q'=\frac{\epsilon_1-\epsilon_2}{\epsilon_1+\epsilon_2}q\hspace{2.5cm}
    q''=\frac{2\epsilon_2}{\epsilon_1+\epsilon_2}q.
\end{equation}
We intend to approach the problem from the semi-classical model of the electric fields. The wavefunction of the electric displacement vector $\Psi_D^q$, due to the real charge, will $reflect$ from the interface and there will be also some $transmission$. Hence, the net wavefunction at $z<0$ is the sum of the wavefunctions due to the real charge and the wavefunction that is $reflected$:
\begin{equation}\label{Eq21}
\Psi_D^{1}(z)=\int v(\mathcal{D}_{1z}) {\bf e}^{i\frac{\int\mathcal{E}_{1z}dz}{\bar\gamma}} d{\mathcal{D}_{1z}} + \scriptr\int v(\mathcal{D}_{1z}){\bf e}^{-i\frac{\int\mathcal{E}_{1z}dz}{\bar\gamma}} d{\mathcal{D}_{1z}},
\end{equation}
where $\scriptr$ denotes the amplitude of reflection back into the material with dielectric constant $\epsilon_1$. On the other hand, the wavefunction at $z>0$ can be written as:
\begin{equation}\label{Eq22}
\Psi_D^2(z)=\scriptt\int v(\mathcal{D}_{2z}) {\bf e}^{i\frac{\int\mathcal{E}_{2z}dz}{\bar\gamma}} d{\mathcal{D}_{2z}},
\end{equation}
where $\scriptt$ denotes the transmission amplitude. To investigate the boundary conditions, we must notice the values of $\Psi_D^{1,2}$ at a distance $|\Delta| \rightarrow0$ on either side of $z=0$. We have seen that $D_ z$ is continuous across the boundary, i.e. $D_{1z}= D_{2z}$. So, we can drop the coefficients while comparing the wavefunctions at both sides close to the boundary. We can also assume that $\mathcal E_{1z}$ and $\mathcal E_{2z}$ vary sufficiently slowly with respect to $z$ near the boundary. This implies that $\int\mathcal E_{1z,2z}dz\approx\mathcal E_{1z,2z}z$. Therefore, from Eq.\eqref{Eq21}, at $z=-\Delta(<0)$ the functional dependence of $\Psi_D^1$ is given as:
\begin{align}\label{Eq23}
\psi_D^{1}(z):={\bf e}^{i\frac{\mathcal{E}_{1z}z}{\bar \gamma}}+\scriptr{\bf e}^{-i\frac{\mathcal{E} _{1z}z}{\bar\gamma}}.
\end{align}
And from Eq.\eqref{Eq22}, we have at $z=\Delta(>0)$:
\begin{equation}\label{Eq24}
\psi_D^2(z):=\scriptt{\bf e}^{i\frac{\mathcal{E}_{2z} z}{\bar\gamma}}.
\end{equation}
The boundary conditions at $z=0$ are:
\begin{subequations}
    \begin{align}
\left(\psi_D^{1}\right)_{z=0^-}&=\left(\psi_D^{2}\right)_{z=0^+}\label{Eq25a}\\
\left(\frac{\partial\psi_D^{1}}{\partial z}\right)_{z=0^-}&= \left(\frac{\partial\psi_D^{2}}{\partial z}\right)_{z=0^+}\label{Eq25b}
    \end{align}
\end{subequations}
The first condition yields:
\begin{align}\label{Eq26}
    1+\scriptr &= \scriptt.
\end{align}
The second condition implies:
\begin{align}\label{Eq27}
\left(\frac{i}{\bar\gamma}\mathcal{E}_{1z}{\bf e}^{\frac {i}{\bar\gamma}\mathcal{E}_{1z}z}-\scriptr\cdot\frac{i}{\bar\gamma}\mathcal{E}_{1z}{\bf e}^{-\frac{i}{\bar\gamma}\mathcal{E}_{1z}z}\right)_{z=0^-}&=\scriptt \left(\frac{i}{\bar\gamma}\mathcal{E}_{2z}{\bf e}^{\frac {i}{\bar\gamma}\mathcal{E}_{2z}z}\right)_{z=0^+}\nonumber \\
\implies \mathcal{E}_{1z}-\scriptr\cdot\mathcal{E}_{1z}&=\scriptt\mathcal{E}_{2z}.
\end{align}
In Eq.\eqref{Eq27}, $-\scriptr\cdot\mathcal{E}_{1z}$ and $\scriptt\mathcal{E}_{2z}$ are reflected and transmitted components of the electric field, respectively. From Eq.\eqref{Eq26} and Eq.\eqref{Eq27}, the reflection amplitude $\scriptr$ at the interface can be calculated as:
\begin{align}\label{Eq28}
    \frac{\mathcal{D}_{1z}}{\epsilon_1}-\scriptr\cdot \frac{\mathcal{D}_{1z}}{\epsilon_1}&=(1+\scriptr)\frac{\mathcal{D}_{2z}}{\epsilon_2}\nonumber\\
\implies\frac{1}{\epsilon_1}-\frac{\scriptr}{\epsilon_1}&=\frac{1}{\epsilon_2}+ \frac{\scriptr}{\epsilon_2}\nonumber\\
\implies\left(\frac{1}{\epsilon_1}-\frac{1}{\epsilon_2}\right)&=\scriptr\left(\frac{1}{\epsilon_1}+\frac{1}{\epsilon_2}\right)\nonumber\\
\implies \scriptr&=-\frac{\epsilon_1-\epsilon_2}{\epsilon_1+\epsilon_2}.
\end{align}
The corresponding transmission amplitude $\scriptt$ can be calculated as:
\begin{equation}\label{Eq29}
 \scriptt = 1+\scriptr=\frac{2\epsilon_2}{\epsilon_1+\epsilon_2}.
\end{equation}
It is worth pointing out that $\scriptr$ and $\scriptt$ are the amplitudes for the wavefunctions of the fields at the interface and are not the coefficients of the image charges required to solve the problem. They scale $\psi_D^{1,2}$ by a constant. In the Hilbert space, the resulting states do not represent any new state. Using the reflected and transmitted wavefunctions in Eq.\eqref{Eq4} and Eq.\eqref{Eq6} does not reproduce the values of the image charges, since the eigenvalue equation does not allow that. These scaled wavefunctions are the results of the boundary conditions and the bound charge present at the interface. 

However, non-vanishing values of these amplitudes suggest that the reflected classical electric field at $z<0$ could be assumed to be due to an $image$ charge $q'$ located at the mirror image position $+z_0$, and the transmitted classical electric field at $z>0$ could be assumed to be due to an $image$ charge $q''$ that is located exactly at the position coincident with the original charge. Referring to Eq.\eqref{Eq27}, we could identify that the reflected electric field $\mathcal{E}_{1z}^{q'}$ due to $q'$ is:
\begin{equation}\label{Eq30}
    \mathcal{E}_{1z}^{q'}=- r\mathcal{E}_{1z}^{q}\implies q'=\frac{\epsilon_1-\epsilon_2}{\epsilon_1+\epsilon_2}q.
\end{equation}
From the same equation, the transmitted electric field $\mathcal{E}_{1z}^{q''}$ due to $q''$ is:
\begin{equation}\label{Eq31}
    \mathcal{E}_{1z}^{q''}=t\mathcal{E}_{2z}^{q}\implies q''=\frac{2\epsilon_2}{\epsilon_1+\epsilon_2}q.
\end{equation}

\subsection{Infinite grounded conducting plane}\label{SubSec5.1}
Let us discuss the relevance of this concept in the context of the infinite grounded conducting plane image problem which can be thought of as the limiting case of the previous problem, in which $\epsilon_2\rightarrow\infty$. In this case, Eq.\eqref{Eq30} shows that one needs to assume an image charge $-q$ inside the conductor. The charge $q''$ is non-zero, however, its contribution to electric potential is vanishing as explained in~\cite{greiner2012classical}. 

We observe an important connection between $\psi_{E_z}$ and potential $\Phi$ in this problem. Since the induced charge density on the surface can be expressed by a delta function, therefore, Eq.\eqref{Eq6} dictates that $\psi_{E_z}=0$ on $z=0$. Classically, in this problem, we
have $\nabla^2\Phi=0$ (Laplace's equation), along with the boundary condition $\Phi=0$ on the conductor. This is exactly parallel to ${\partial^2}|\psi_{Ez}|^2/{\partial z^2}=0$ (Eq.\eqref{Eq9}), with the boundary condition $\psi_{Ez}=0$ (implying $|\psi_{Ez}|^2=0$) at $z=0$. Therefore, we conclude that in this problem, $\Phi \propto|\psi_{Ez}|^2$. This argument is also applicable for the grounded conducting sphere image problem~\cite{griffiths2005introduction} where one is asked to calculate the image charge and location when a real charge is placed in front of a grounded conducting sphere. 

It may be noted that the superposition of $\psi_{Ez}$ must be done in the quantum mechanical sense taking into account the phase, as demonstrated in~\cite{bhattacharya2021demystifying}. However, in the square of the modulus of $\psi_{Ez}$, relevant in several electrostatics problems, the sensitivity to the phase is washed out. No wonder that electrostatic potential obeys the classical superposition principle, where one adds up the potentials algebraically, without any reference to phase.

\section{Semi-classical limit of Gauss's law or Poisson's equation}\label{Section6}
At this point, it is perhaps good to note the difference between the two frameworks that discuss the quantum theory of electrostatics and the possible expression of Gauss's law in this limit. First of all, we are not interested in the derivation of the classical version of Gauss's law or its higher-order corrections in the non-relativistic limit of QED which have been addressed~\cite{dybalski2016non}. We are specifically interested in the form of Gauss's law (or wave equation) in which the non-travelling wavefunctions (or non-travelling wavefunctions (or `electrostatic coherent state—a notion which involves (non-dynamical) longitudinal photons') of the electric field can be calculated from a given source charge distribution. The framework presented in~\cite{kay2022quantum} assumes that Gauss's law should hold in quantum theory as $(\nabla\cdot\vec{\mathcal E})\Psi=\rho\Psi$, or as its expectation value thereof (see the discussion at pp.3-4). 
In this equation, the electric field is an operator.
On the other hand, the framework on electrostatic field theory, presented in~\cite{bhattacharya2021demystifying} and extended in the present work, the electrostatic field is taken as a conservative vector field which can have a semi-classical nature in a limit. We called this framework semi-classical, because the main features of the quantum wavefunction are illuminated by classical physics. This framework seems to be consistent with classical physics. This framework provides another form of Gauss's law, given in Eq.\eqref{EqQuantumGauss}. Most likely, the difference arises due to the fact that the latter is a semi-classical model, as opposed to the former which derives from quantum field theory. However, we comment that to check consistency with the classical results, the semi-classical model should be a better starting point. In this case, we must deal with a vector of wavefunctions representing the three components of the electrostatic field. Eq.\eqref{EqQuantumGauss}, when solved, would give the wavefunctions representing three different components of the electrostatic field due to a source charge distribution $\rho$, just like solving the Gauss's law (or Poisson's equation) in electrostatics can be used to evaluate the electric field in a problem.

In principle, there can be a spherical symmetry in the source charge distribution, for which Eq.\eqref{Eq5a}, Eq.\eqref{Eq5b}, and Eq.\eqref{Eq5c} will fuse into the basic Eq.\eqref{Eq3}. In that case, the semi-classical version of Gauss's law (or Poisson's equation) takes a simpler form:
\begin{align}\label{Eq33}
    -i\bar\gamma\frac{1}{\psi_E}\nabla\psi_E&=-i\bar\gamma\nabla{(\ln\psi_E)}=\vec{\mathcal{E}}\nonumber\\
\implies -i\bar\gamma\nabla^2(\ln\psi_E)&=\nabla\cdot\vec{\mathcal{E}}=\frac{\rho}{\epsilon_0}\nonumber\\
\implies\nabla^2(\ln\psi_E)&=i\frac{\rho}{\bar\gamma\epsilon_0}.
\end{align}
This equation should be interpreted as a method to find $\psi_E$, the wavefunction of an electrostatic field due to a spherically symmetric $\rho$. As expected, symmetry leads to considerable simplification of the problem. We comment that the eigenvalue equation Eq.\eqref{Eq3} can be thought of as the analogue of the gradient equation: $ \vec{\mathcal{E}}=-\nabla\Phi$, because when the divergence operator is applied to it, we get the Gauss's law (or the Poisson's equation).

\section{Summary and Discussions}\label{Section7}
In this paper, we gave an explanation of the quantization of electric charge without requiring the existence of magnetic monopole, based on a semi-classical model of curl-free vector fields~\cite{bhattacharya2021demystifying} developed in the context of resolving the nonlocality problem of the Aharonov-Bohm effect~\cite{aharonov1959significance}. Through this exercise, we resolved an open problem in physics that remained a mystery for about the last hundred years. Apart from charge quantization, the semi-classical model of static conservative fields has been found to resolve the nonlocality problem of the Aharonov-Bohm effect (which was also an open problem since 1959), Aharonov-Casher effect and He-McKellar-Wilkens effect; for deriving the magnetic flux quantum, etc. All these effects, except for the electrostatic Aharonov-Bohm effect have been experimentally validated. The reason why the electrostatic Aharonov-Bohm effect could not be observed is usually attributed to the difficulty in reducing the electric field to zero in the experimental setup, in contradiction with the preamble of the original thought experiment. On the other hand, the consistency of this model with known results from classical physics further emphasises its validity. The plane waves of the electrostatic fields are found to be similar to the plane waves used in geometrical optics. The author was thrilled to find that the electrostatic potential and the modulus square of the wavefunctions representing an electrostatic field could be related if they are subjected to the same boundary condition. Its implication for the difference between the quantum and classical superposition principles (discussed at the end of subsection~\ref{SubSec5.1}) must be appreciated. It should be possible to deduce the equation corresponding to Eq.\eqref{Eq19} in the context of the passage of a magnetic field line between two media with different magnetic permeabilities, based on a semi-classical model on magnetostatics. 

In the literature, we found another paper on the quantum physical model of electrostatics proposed by Kay~\cite{kay2022quantum}, based on his previous paper~\cite{kay1998decoherence}. These works are based on quantum electrodynamics and describe the wavefunctions of electric fields as coherent states. The author talks about two frameworks in which Gauss's law in electrostatics holds as an operator equation ($(\nabla\cdot\vec{\mathcal{E}})\Psi=\rho\Psi$) or its expectation. However, it is not clear if this model can be used to achieve the tasks we performed.

At this point, it may be interesting to note the relationship between the semi-classical model with QED, in which electrostatic force arises due to the interaction between bodies mediated by the virtual photons that exist for a very short time scale determined by the uncertainty principle~\cite{zee2010quantum}. Though QED is undoubtedly the most successful theory, it is perhaps safe to say that the time-invariant picture of classical electrostatics is not intuitive from the dynamical description of the fields in quantum electrodynamics [see the discussion after Eq.(1) in the introduction of~\cite{bhattacharya2021demystifying}]. In addition, quantization of electromagnetic Hamiltonian by treating it effectively as a harmonic oscillator may be difficult in a frame where there is only an electrostatic field, but no magnetic field. This is perhaps the key difference between QED and the current formulation. To reproduce the present model from QED, one needs to come out of the traditional photon picture, by transforming to a reference frame where only an electrostatic field is present. It may not be possible to directly take a limit and reproduce the model. This is because the variational principle $\delta\int_{P_1}^{P_2} \mathcal{E}ds=0$~\cite{bhattacharya2022unexplored} which is the basis of the semi-classical model, is an independent component that was not known explicitly to be a part of the standard electrostatics framework. This principle looks like Fermat’s principle, but cannot be derived from it. The standard theory of QED does not explicitly have this piece of information. However, this relationship is definitely worth exploring. It is possible that combining this model with QED will lead to a more complete theory.

Finally, we comment that the semi-classical model lacks the important aspect of spin. This aspect arises in the QED based-model~\cite{kay2022quantum} naturally. We did not need it for the problems discussed in this paper. A more complete model, however, can be formulated if spin can be taken into account. It may be possible to accomplish this using the square-root operator, as shown in the context of light rays in optics~\cite{eichmann1971quasi}. 

\section{Acknowledgement}
The author acknowledges the anonymous reviewers for providing useful suggestions; Prof. Anwesh Mazumdar and Prof. Sudipto Roy for their support and encouragement.
\section{Data availability statement}
No new data were created or analysed in this study.
\section{References}
\bibliographystyle{unsrt}
\bibliography{QM-image}


\end{document}